\documentclass{aa}
\usepackage{psfig}

\def\la{\;
\raise0.3ex\hbox{$<$\kern-0.75em\raise-1.1ex\hbox{$\sim$}}\; }
\def\ga{\;
\raise0.3ex\hbox{$>$\kern-0.75em\raise-1.1ex\hbox{$\sim$}}\; }

\newcommand{\zabs}{$z_{\rm abs}\,$}
\newcommand{\zem}{$z_{\rm em}\,$}

\newcommand{\kms}{km~s$^{-1}\,$}
\newcommand{\cm}{cm$^{-2}\,$}
\newcommand{\cmm}{cm$^{-3}\,$}

\begin{document}

\title{The broad \ion{O}{vi} absorber at \textit{z}\,=\,3.02 toward 
CTQ~325\thanks{Based 
on observations obtained at the VLT Kueyen telescope (ESO, Paranal, Chile),
the ESO programme 69.A-0123.
}
}

\author{
S. A. Levshakov\inst{1}
\and
S. D'Odorico\inst{2}
\and
I. I. Agafonova\inst{1}
\and
M. Dessauges-Zavadsky\inst{3}
}

\offprints{S. A. Levshakov}

\institute{Department of Theoretical Astrophysics,
Ioffe Physico-Technical Institute, 194021 St.Petersburg, Russia
\and
European Southern Observatory, Karl-Schwarzschild-Strasse 2,
D-85748 Garching bei M\"unchen, Germany
\and
Observatoire de Gen\`eve, CH-1290 Sauverny, Switzerland
}
\date{Received 00  / Accepted 00 }

\abstract{We report on the multiphase
absorption-line system detected at a redshift \zabs=3.021 
in the spectrum of the quasi-stellar object CTQ~325 (\zem = 3.212). 
The system is displaced by $\simeq 14,000$ \kms\, to the blue of the
systemic velocity defined by the center of the symmetric \ion{O}{i} quasar emission line.
It consists of shallow and broad 
($\sim700$ \kms\, FWHM) absorption lines of \ion{H}{i} $\lambda1215$,
\ion{C}{iv} $\lambda\lambda1548, 1550$, and \ion{O}{vi} $\lambda\lambda1032, 1038$,
produced by collisionally ionized gas and of
narrow absorption lines (FWHM $< 20$ \kms) of Ly$_{\alpha, \beta, \ldots, 6}$,
\ion{Si}{iii} $\lambda1206$, and \ion{C}{iii} $\lambda977$ produced
by photoionized gas.
Possible origins 
of this system are discussed.
}

\authorrunning{S. A. Levshakov et al.}
\titlerunning{The broad \ion{O}{vi} 
absorber at $z=3.02$ toward CTQ~325}
\maketitle

\section{Introduction}

While performing a study of the high-resolution spectrum of 
the distant quasar CTQ~325 from the Cal\'an-Tololo Survey (Maza et al. 1996),
we experienced problems in the continuum
placement and the subsequent analysis of the normalized data for the 
regions $\lambda\lambda$ = 4100-4200 \AA,  4850-4950 \AA\,
and  6200-6280 \AA. The original
spectrum, shown in Fig.~1, exhibits bumps and shallow troughs in these regions
which could hardly be expected from the QSO continuum itself.
Can these irregularities be attributed to
shallow broad absorption lines or are they artifacts resulting from inappropriate 
data reduction? 

Calibration flux errors can be excluded in our case since 
these features are present in all three unnormalized spectra 
we have for this range 
of exposure time of 4100s each. 
The broad features are located in the middle of echelle orders, so
they cannot be explained by a bad merging of the orders as well.
The spectrum of other quasar CTQ~298 taken during the same set
of observations with the UVES/VLT does not show any peculiarities
in these spectral ranges. 
In a composite QSO spectrum (Telfer et al. 2002) there are no
weak emission-line features which could mimic bumps and troughs
in the Ly$\alpha$ forest similar to that observed in CTQ~325.
Thus we conclude that the broad absorption features we see in CTQ~325 are real. 

The identification of these lines is straightforward: if
the broad absorption at $\lambda \simeq$ 6225-6230 \AA\, 
is the unresolved doublet
\ion{C}{iv} $\lambda\lambda1548,1551$ at \zabs = 3.021, 
then the absorptions at $\lambda \simeq 4172$, 4149, and
4888 \AA\, are, respectively, \ion{O}{vi} $\lambda 1032$, 
\ion{O}{vi} $\lambda 1038$, and Ly$\alpha$
(the expected positions of \ion{N}{v} $\lambda\lambda 1238, 1242$
are blended with the blue wing of the damped Ly$\alpha$ at
\zabs = 3.118).

\begin{figure*}
\vspace{0.0cm}
\hspace{0.0cm}\psfig{figure=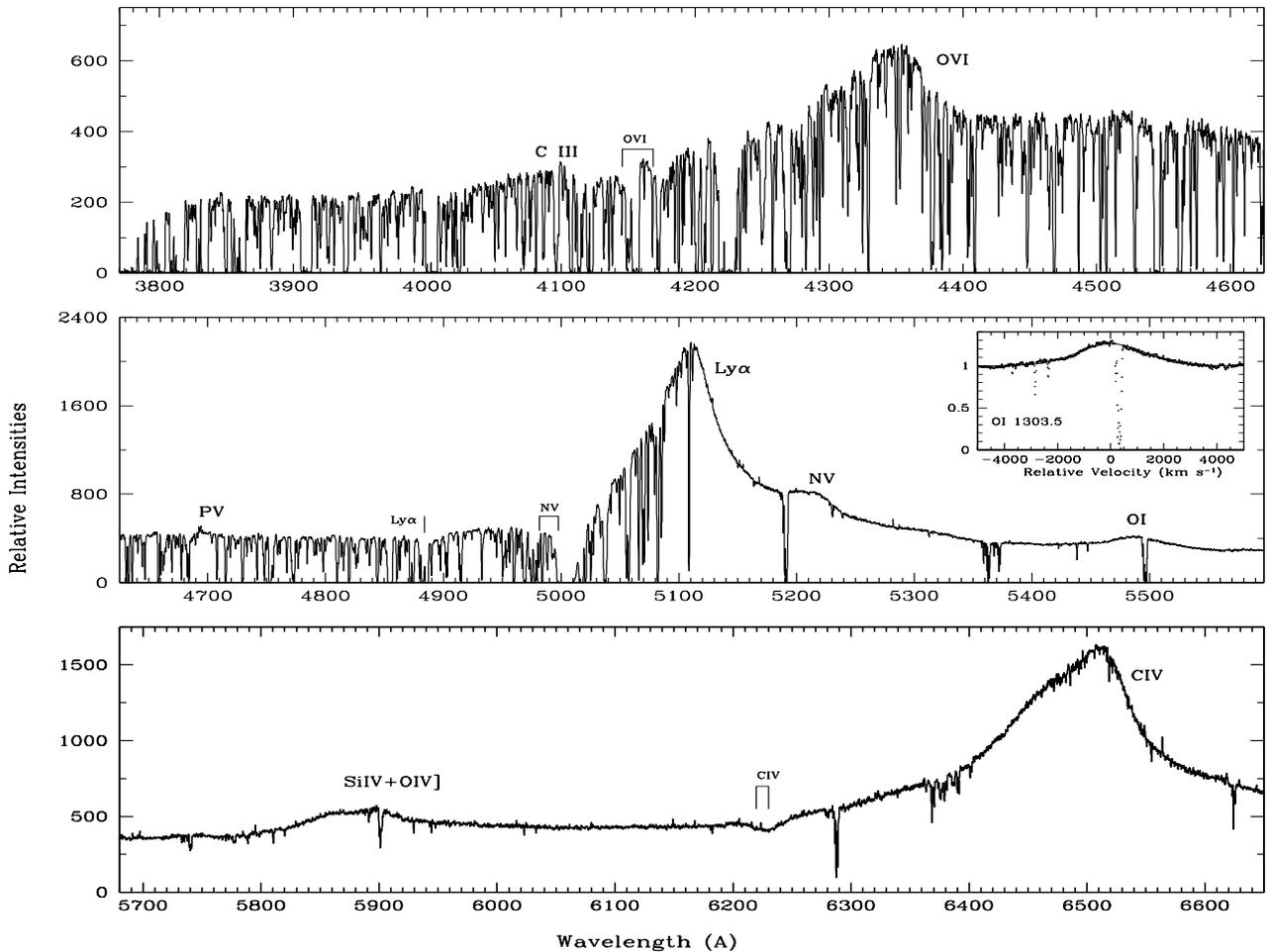,height=13.0cm,width=18.0cm}
\vspace{-0.5cm}
\caption[]{
VLT/UVES spectra of CTQ~325 as a function of vacuum heliocentric wavelength.
The spectral resolution is $\simeq 6.5$ \kms\, (FWHM). The identification of
the emission lines are indicated near each spectrum. A smooth fit by a Gaussian
(solid curve) to the normalized \ion{O}{i} $\lambda1303.5$ emission profile (dots),
shown in the inset in the middle panel, provides the QSO redshift \zem = 
$3.21196\pm0.00010$ (2 $\sigma$). The peaks of the asymmetric Ly$\alpha$, 
\ion{C}{iv} $\lambda1549.1$ and a weak
\ion{P}{v} $\lambda 1118$ emission lines are blueshifted by $\sim 570$ \kms\,
with respect to this \ion{O}{i} line 
giving $z'_{\rm em} = 3.20$ in agreement with 
low-resolution observations by Lopez et al. (2001).
Vertical marks indicate the broad absorption lines at \zabs = 3.0212.
}
\label{fig1}
\end{figure*}

Similar shallow (the central optical depth $\tau_0 < 1$) broad 
(FWHM $> 100$ \kms) absorption line systems 
can be found in other QSOs.
For example, 
in the spectrum of PG 2302+029 (\zem=1.052)
a very broad (FWHM $> 3000$ \kms) absorption complex 
consisting of unresolved doublets of \ion{C}{iv}, \ion{N}{v}, and  \ion{O}{vi}
(Ly$\alpha$ is undetectable) 
was found at a redshift $\sim0.695$ that corresponds to the blueshifted 
`ejection' velocity of
$\sim56,000$ \kms (Jannuzi et al. 1996).
A weak \ion{C}{iv} trough with FWHM $\sim 800$ \kms\, 
at the blueshifted velocity $\sim4000$ \kms\,
(\zabs = 4.045)
is present in the spectrum of FIRST 0747+2739 --- a quasar with a redshift of 4.11
(Richards et al. 2002). 
Besides, in the published spectrum 
the shallow and broad absorption features at the expected positions of
the Ly$\alpha$ and \ion{N}{v} $\lambda1242$ lines
can be found as well (see Fig.~1 in Richards et al.).
An absorption system at $z=2.146$
with broad and shallow \ion{C}{iv} and \ion{N}{v} doublets 
(FWHM $\ga 1000$ \kms)
and weak Ly$\alpha$ was identified in the
spectrum of Tol 1038--2712 (\zem = 2.331) which belongs to the famous
`Tololo pair' (e.g., Dinshaw \& Impey 1996; Srianand \& Petitjean 2001).

The interpretation of these shallow and broad systems is not unique. 
Highly ionized gas can reside inside the quasar host galaxy, be ejected
from the galaxy, or trace the intergalactic medium (IGM).
For instance, the absorber in PG 2302+029 is intrinsic since
the comparison of spectra taken for this QSO in 1994 and 1998
revealed that `\ion{O}{vi} and \ion{N}{v} dramatically weakened to become
unmeasurable' (Sabra et al. 2003).
On the other hand, the above mentioned Tololo absorber is probably an example
of the intervening system (a putative supercluster or filament)
since a damped Ly$\alpha$ system [$\log N$(\ion{H}{i}) $\sim 19.7$, from Srianand \&
Petitjean 2001] 
at the same redshift $z=2.14$ is seen
in the companion quasar Tol 1037--2704 (\zem = 2.207) separated by $\sim 4$ Mpc from
the line of sight toward Tol 1038--2712. 
In general, the interpretation of quasar absorption-line systems is not such
straightforward and requires accounting for all available data.

In this paper, we present the quantitative analysis of the \zabs = 3.0212
system toward CTQ~325. Possible explanations of its nature are discussed in Sect.~4.

\begin{figure}
\vspace{0.0cm}
\hspace{0.0cm}\psfig{figure=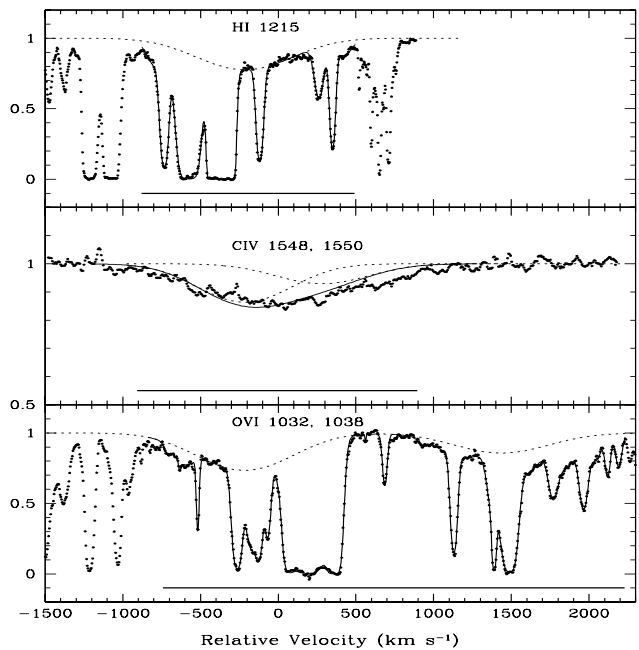,height=6.0cm,width=12.0cm}
\vspace{2.5cm}
\caption[]{
Observed absorption lines in the spectrum of CTQ~325 (dots)
and over-plotted synthetic profiles (solid curves) calculated from the
joint analysis of the shallow broad and narrow components.
The zero radial velocity
is fixed at \zabs = 3.0212. Bold horizontal lines mark pixels included in
the optimization procedure. In addition, synthetic profiles of broad Ly$\alpha$,
\ion{C}{iv} $\lambda\lambda 1548, 1550$, and
\ion{O}{vi} $\lambda\lambda 1032, 1038$ are shown separately
(dashed curves).
For each doublet, the velocity scale refers to the first component.
}
\label{fig2}
\end{figure}

\section{Observations}

The quasar CTQ~325 
($\alpha = 13^{\rm h}42^{\rm m}58.9^{\rm s}$,
$\delta = -13^\circ55'59.9''$, J2000; $m_{B} = 18.3$; Lopez et al. 2001)
was observed with the VLT/UVES over the nights 16-18 March, 2002.
The spectrum covers the range between 3750 and 9900 \AA,
with wavelength gaps between 5597--5677 \AA\, and 7940--8077 \AA.
The resulting spectral resolution is between 6.3 and 6.8 \kms\, FWHM, 
and the S/N ratio
per pixel $\simeq$ 40-50 ($\lambda\lambda \simeq$ 3800-4700 \AA) 
and S/N $> 50$ ($\lambda\lambda \simeq$ 4700-6600 \AA).
The data reduction was performed using the ECHELLE context routines implemented
in the ESO MIDAS package. 
The details of the reduction procedure is outlined in Molaro et al. (2000).
The uncertainties in the continuum placement are not larger than 5\% due to high
S/N and high spectral resolution.

\section{Analysis}

{\it Emission-line redshift.} One of the noticeable features 
of the spectrum of CTQ~325
is that the blue side wing of the \ion{C}{iv} emission line, 
which is not affected by numerous absorption
lines, has 
a longer tale than the redward side (Fig.~1). This blue-side
asymmetry is also seen in the profiles 
of Ly$\alpha$
and \ion{O}{vi}, whereas the neutral oxygen line \ion{O}{i} $\lambda 1303.5$ is
symmetric, providing us with the accurate measurement 
of the emission redshift of the
QSO, \zem = $3.21196\pm0.00010$ (2 $\sigma$).

From earlier studies (e.g., Gaskell 1982; Espey et al. 1989; Tytler \& Fan 1992)
it was found that the mean blueshift of Ly$\alpha$, \ion{C}{iv}, and \ion{N}{v} with
respect to \ion{O}{i} and \ion{Mg}{ii} is about 600 \kms.
The same order of magnitude blueshift of $\sim 570$ \kms\, was measured in our case
between the center of the \ion{O}{i} line and the peaks of the Ly$\alpha$, 
\ion{C}{iv}, and \ion{P}{v} lines. 
This could indicate that the redshift of CTQ~325 is slightly
higher than $z'_{\rm em} = 3.20$ previously deduced from low-resolution data
(FWHM = 5.2 \AA) by Lopez et al. (2001).

\begin{figure}
\vspace{0.0cm}
\hspace{0.0cm}\psfig{figure=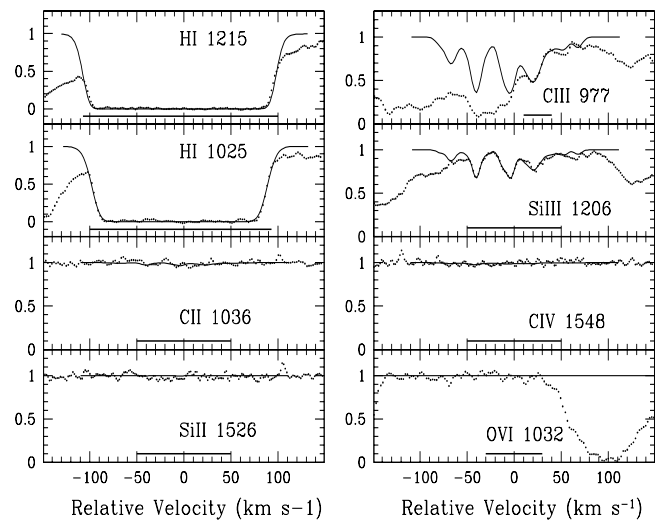,height=11.5cm,width=10.0cm}
\vspace{-5.0cm}
\caption[]{
Hydrogen and metal lines associated with the
\zabs = 3.0192 low-ionized absorber toward CTQ~325 (dots).
Smooth lines are
the synthetic spectra convolved with the corresponding point-spread
functions.
Bold horizontal lines mark pixels included
in the MCI procedure. 
Continuum windows at the positions of \ion{C}{ii} $\lambda 1036$,
\ion{Si}{ii} $\lambda 1526$, \ion{C}{iv} $\lambda 1548$,
and \ion{O}{vi} $\lambda 1032$ are shown
as well.
The normalized $\chi^2_{\rm min} = 0.92$ (the number of degrees of
freedom $\nu = 400$). 
}
\label{fig3}
\end{figure}

\medskip\noindent
{\it The \ion{O}{vi} absorber at \zabs = 3.0212.} The analysis of the 
broad complex is illustrated in Fig.~2. 
We found that all lines can be well fitted by a simple Gaussian with the broadening
$b$-parameter of $412\pm20$ \kms, and column densities
$N$(\ion{H}{i}) = $(1.38\pm0.07)\times10^{14}$ \cm,
$N$(\ion{C}{iv}) = $(1.37\pm0.07)\times10^{14}$ \cm, and
$N$(\ion{O}{vi}) = $(6.2\pm0.3)\times10^{14}$ \cm.
Since the shallow 
profiles of the Ly$\alpha$ and \ion{O}{vi}
lines are affected by the narrow forest lines, 
we included in the fitting
procedure the whole regions marked in Fig.~2 by the bold horizontal lines.
All three regions with the broad \ion{H}{i} $\lambda 1215$, 
\ion{C}{iv} $\lambda\lambda 1548, 1550$,
and \ion{O}{vi} $\lambda\lambda 1032, 1038$ were analyzed simultaneously.
The corresponding equivalent widths are as follows (rest frame values):
$W_\lambda$(\ion{H}{i}$_{1215}$) $\simeq 190$ m\AA, 
$W_\lambda$(\ion{C}{iv}$_{1548+1550}$) $\simeq 210$ m\AA, 
$W_\lambda$(\ion{O}{vi}$_{1032}$) $\simeq 190$ m\AA, and
$W_\lambda$(\ion{O}{vi}$_{1038}$) $\simeq 95$ m\AA, 
and the central optical depths are $\tau_0 = 0.25$, 0.15, 0.31, and 0.15,
respectively.

The estimated column density ratios 
$N$(\ion{C}{iv})/$N$(\ion{H}{i}) $\sim 1$ and 
$N$(\ion{O}{vi})/$N$(\ion{H}{i}) $\sim 4.5$ indicate that the gas
is collisionally ionized and the kinetic temperature $\log T > 5$ 
(see Figs.~1 and 2 in Levshakov et al. 2003a). 
If we assume that the metallicity pattern is solar,
then according to Sutherland \& Dopita (1993) the optimal value
for the kinetic temperature is $\log T = 5.30$.
At this temperature we find the total column densities of
hydrogen, carbon, and oxygen being
$N$(H) $\simeq 4.3\times10^{19}$ \cm,
$N$(C) $\simeq 1.2\times10^{16}$ \cm, and 
$N$(O) $\simeq 3.0\times10^{16}$ \cm, i.e. the metal content in this
absorber is slightly over solar, [C/H] $\simeq 0.05$ and 
[O/H] $\simeq 0.15$ 
\footnote{
[X/Y] $\equiv$ log (X/Y) -- log (X/Y)$_\odot$ .
Throughout the text photospheric solar abundances
for C and O are taken from Allende Prieto et al. (2001, 2002),
and for Si from Holweger (2001).
}.

The narrow absorption lines beneath the broad Ly$\alpha$ 
in Fig.~2 at $\Delta v \simeq$
--519, --365, --150, and 95 \kms\, 
are the Ly$\alpha$ lines which have corresponding
Ly$\beta$ and higher order Lyman series lines up to Ly$_6$. 
The strongest of these
absorbers, that at
$\Delta v \simeq -150$ \kms, shows 
metal absorption in the \ion{Si}{iii} $\lambda 1206$
and partially blended \ion{C}{iii} $\lambda 977$ lines and 
line-free continuum
at the expected positions of the
\ion{C}{ii} $\lambda 1334$, \ion{Si}{ii} $\lambda 1526$,
\ion{C}{iv} $\lambda 1548$, and \ion{O}{vi} $\lambda 1032$ lines (Fig.~3).

\begin{figure}
\vspace{0.0cm}
\hspace{-0.5cm}\psfig{figure=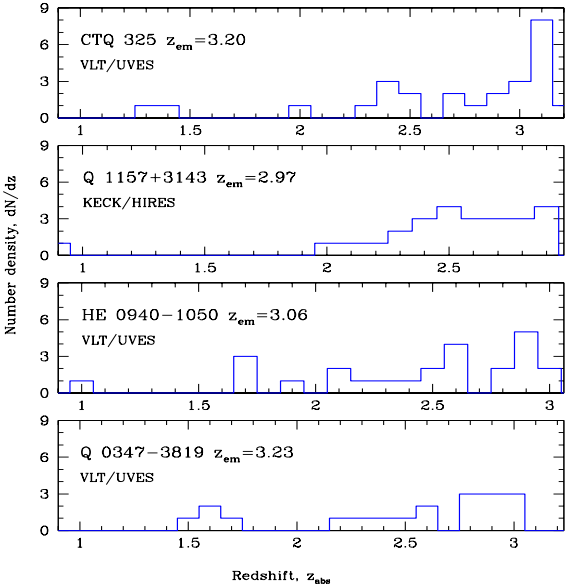,height=12.0cm,width=12.0cm}
\vspace{-4.5cm}
\caption[]{Number density distributions of metal absorbers toward
CTQ~325 (this paper), Q~1157+3143 and HE~0940--1050 (Levshakov et al. 2003c),
and Q~0347--3819 (Levshakov et al. 2002b).
QSO spectra with similar quality observations are used.
The adopted bin size is $\Delta z = 0.1$.
}
\label{fig4}
\end{figure}

We have analyzed this metal system with the Monte Carlo Inversion (MCI) procedure
(Levshakov et al. 2000, 2002a) with three different UV ionizing backgrounds:
the power law $J_\nu \propto \nu^{-1.5}$, the 
Mathews-Ferland spectrum
(Mathews \& Ferland 1987, hereafter MF), and the Haardt-Madau spectrum
(Haardt \& Madau 1996, hereafter HM). 
Since hydrogen lines in this system are saturated, we were able
to estimate only the lower limit of $N$(\ion{H}{i}) $\ga 2\times10^{16}$ \cm.
Analyzing simultaneously all available metal lines (including those presented
only as upper limits), we obtained the mean ionization parameter
$U_0 \simeq 1.3\times10^{-3}$, 
total hydrogen column density $N$(H) $\ga 10^{18}$ \cm\, and metal content
[C/H] $\la -1.0$, [Si/H] $\la -0.9$ for the MF and $\nu^{-1.5}$ spectra, and
$U_0 \simeq 2.4\times10^{-3}$, 
$N$(H) $\ga 5\times10^{18}$ \cm\, and
[C/H] $\la -1.5$, [Si/H] $\la -1.5$ for the HM spectrum.
The estimated column densities for particular ions are as follows: 
$N$(\ion{C}{ii}) $< 6.0\times10^{12}$ \cm,
$N$(\ion{Si}{ii}) $< 6.0\times10^{11}$ \cm,
$N$(\ion{Si}{iii}) $= (3.5\pm0.4)\times10^{12}$ \cm, 
$N$(\ion{C}{iii}) $\sim 3\times10^{13}$ \cm\, (uncertain because of blending), and
$N$(\ion{C}{iv}) $< 2.0\times10^{12}$ \cm.

Taking the metagalactic value of the
mean specific intensity at the hydrogen Lyman edge
$J_{912}$($z$=3) $ = 0.4\times10^{-21}$ ergs cm$^{-2}$ s$^{-1}$ Hz$^{-1}$ sr$^{-1}$
from HM,
we calculate the mean number density $n_0 \simeq 3.5\times10^{-2}$ \cmm\,
and the linear size $L \ga 10$ pc (MF and $\nu^{-1.5}$), and
$n_0 \simeq 10^{-2}$ \cmm, $L \ga 150$ pc (HM).   
The mean kinetic temperature is estimated as $(1.5-2)\times10^4$ K.
From the non-detection of \ion{C}{ii} and
\ion{Si}{ii} lines we can also deduce the upper limit of
$N$(\ion{H}{i}) $< 10^{18}$ \cm, since otherwise the self-shielding, 
important at $N$(\ion{H}{i}) $> 3\times10^{17}$ \cm, would shift
the ionization state toward these
low-ionization transitions.
Thus, in any case the size of the absorber at $\Delta v = -150$ \kms\,
is not larger than 1 kpc.
The gas density will be substantially higher and the linear size smaller
if the absorber is located close to the QSO where the intensity of
the local ionizing background is enhanced by the direct QSO radiation.

Unfortunately, other narrow Ly$\alpha$-line absorbers
have blends at the expected positions of metals which makes impossible
any quantitative conclusions about their physics.

\begin{table}
\centering
\caption{Metal absorbers toward CTQ~325}
\label{tbl-1}
\begin{tabular}{cl}
\hline
\noalign{\smallskip}
\zabs & Species identified \\
\hline
\noalign{\smallskip}
1.3975 & \ion{Mg}{ii}$_{2796, 2803}$, \ion{Al}{iii}$_{1854, 1862}$ \\
1.4031 & \ion{Mg}{ii}$_{2796, 2803}$, \ion{Al}{ii}$_{1670}$, \ion{Al}{iii}$_{1854, 1862}$ \\
2.0902 & \ion{C}{iv}$_{1548, 1550}$ \\
2.3785 & Ly$_\alpha$, \ion{C}{iv}$_{1548, 1550}$ \\
2.4315 & Ly$_\alpha$, \ion{Si}{iii}$_{1206}$, \ion{C}{iv}$_{1548, 1550}$, \ion{Si}{iv}$_{1393}$ \\
2.4615 & Ly$_\alpha$, \ion{C}{iv}$_{1548, 1550}$ \\
2.4775 & Ly$_\alpha$, \ion{Si}{iii}$_{1206}$, \ion{C}{iv}$_{1550}$ \\
2.5025 & Ly$_\alpha$, \ion{C}{iv}$_{1548, 1550}$ \\
2.5126 & Ly$_\alpha$, \ion{C}{iv}$_{1548, 1550}$ \\
2.7392 & Ly$_{\alpha, \beta}$, \ion{C}{iv}$_{1548, 1550}$ \\
2.7531 & Ly$_{\alpha, \beta}$, \ion{C}{iv}$_{1548, 1550}$ \\
2.8903 & Ly$_{\alpha, \beta, \gamma}$, \ion{C}{iv}$_{1548, 1550}$ \\
2.9098 & Ly$_{\alpha, \beta}$, \ion{O}{vi}$_{1031}$ \\
2.9933 & Ly$_{\alpha, \beta, \gamma}$, \ion{C}{iv}$_{1548, 1550}$ \\
3.0096 & Ly$_{\alpha, \beta, \gamma, \delta}$, \ion{C}{iii}$_{977}$, \ion{C}{iv}$_{1548}$ \\
3.0212 & Ly$_{\alpha}$, \ion{C}{iv}$_{1548, 1550}$, \ion{O}{vi}$_{1031, 1037}$ \\
3.0680 & Ly$_\alpha$, \ion{C}{iv}$_{1548, 1550}$ \\
3.1136 & Ly$_{\alpha, \ldots, 10}$, \ion{C}{iii}$_{977}$, \ion{C}{iv}$_{1548, 1550}$, \ion{N}{v}$_{1238}$ 
	   \ion{O}{vi}$_{1031}$\\
3.1148 & Ly$_{\alpha, \ldots, 9}$, \ion{C}{iii}$_{977}$, \ion{Si}{iii}$_{1206}$,
	 \ion{C}{iv}$_{1548, 1550}$, \ion{Si}{iv}$_{1393, 1402}$, \\ 
	  & \ion{O}{vi}$_{1031}$\\
3.1184 & Ly$_{\alpha, \ldots, 10}$, \ion{N}{i}$_{1199.5, 1200.2}$, \ion{O}{i}$_{950, 976, 988, 1039, 1302}$, \\
       & \ion{Ar}{i}$_{1048}$, \ion{C}{ii}$_{1036, 1334}$, \ion{N}{ii}$_{1083}$, \ion{Mg}{ii}$_{1239}$, 
	  \ion{Al}{ii}$_{1670}$,\\
       &  \ion{Si}{ii}$_{989, 1020, 1190, 1193, 1260, 1304, 1526, 1808}$,  \ion{P}{ii}$_{963, 1152}$,\\ 
       & \ion{P}{ii}$_{1301}$, \ion{S}{ii}$_{1250, 1253, 1259}$, \ion{Fe}{ii}$_{940, 1063, 1081, 1096, 1608}$,\\
       & \ion{Ni}{ii}$_{1317, 1454, 1741}$, \ion{C}{iii}$_{977}$, \ion{Si}{iii}$_{1206}$, 
       \ion{C}{iv}$_{1548, 1550}$ \\
3.1274 & Ly$_{\alpha, \ldots, 7}$, \ion{C}{iii}$_{977}$, \ion{C}{iv}$_{1548, 1550}$, \ion{O}{vi}$_{1031}$ \\
3.1284 & Ly$_{\alpha, \ldots, 7}$, \ion{C}{iii}$_{977}$, \ion{C}{iv}$_{1548, 1550}$, \ion{Si}{iv}$_{1393}$ 
	 \ion{N}{v}$_{1238}$\\
	 & \ion{O}{vi}$_{1031}$ \\
3.1438 & Ly$_{\alpha, \beta, \gamma, \delta}$, \ion{C}{iv}$_{1548}$, \ion{O}{vi}$_{1037}$ \\
3.1701 & Ly$_{\alpha, \beta}$, \ion{O}{vi}$_{1031}$ \\
3.1711 & Ly$_{\alpha, \beta, \gamma}$, \ion{O}{vi}$_{1031}$ \\
3.2024 & Ly$_{\alpha, \beta}$, \ion{N}{v}$_{1238}$, \ion{O}{vi}$_{1031, 1037}$ \\
\noalign{\smallskip}
\hline
\end{tabular}
\end{table}

\section{Discussion}

Equipped with the results obtained in the previous
section we can now consider the possible causes for the
observed absorptions.

The narrow line system is probably physically connected
with the shallow absorber since both the detected ions
(\ion{Si}{iii} and \ion{C}{iii})
and the linear size ($L \sim 100$ pc) are very unusual for
a `stand alone' intervening system: most QSO absorbers exhibit 
\ion{C}{iv} lines and have linear sizes of $L \sim 10s$ kpc. 
Thus what we observe is a metal-poor cool and photoionized cloud(s)
embedded in hot
collisionally ionized gas with solar metal content.
Most plausible origin of this structure can provide
the jet-cloud interaction -- the phenomenon well known
from the radio observations of active galactic nuclei 
(e.g., Carvalho \& O'Dea 2002 and references therein). 
In this scenario, the shock (blast) wave driven by the QSO jet
strikes an intervening cloud embedded in the interstellar medium.
Model calculations and model experiments
of shock-cloud interaction (Klein et al.  1994; Klein et al. 2003) 
show that this process leads to the cloud fragmentation and
produces the velocity dispersion of $\sim 0.1v_{\rm blast}$.
After fragmentation, the cloud material continues to be accelerated
until it is approximately comoving with the surrounding gas in the jet.
In addition, shock compresses the cloud gas which can enhance its
cooling rate.
It is to be noted that typical parameters of the AGN jets are: 
radius $\sim 500$ pc,
length-to-radius ratio $\sim 100:1$, and
expansion velocity $\sim$ 1000s \kms\,
(Carvalho \& O'Dea 2002).

Difference in metallicities between the collisionally ionized gas
and gas in the narrow-line absorber can also be explained within
the frame of the jet-cloud interaction. Namely,
solar and oversolar metallicities are characteristic for the QSO
circumnuclear gas (e.g., Hamann \& Ferland 1999), whereas the metallicity
of the ambient interstellar matter is much lower ($Z < 0.1 Z_\odot$).

The profiles of the QSO emission lines Ly$\alpha$, \ion{C}{iv}, and
\ion{O}{vi} are highly asymmetric indicating strong outflows. The shallow and
broad absorber is blueshifted by $\sim 14,000$ \kms\, from the systemic velocity
(defined by the center of the symmetric \ion{O}{i} quasar emission line)
and lies at the high ejection end of the asymmetric emission lines. 
Whether this is a chance
coincidence or the absorber is indeed located very close to the QSO emission line
region can be tested in monitoring this quasar over the course of a few years: 
variability of the absorption profiles would testify the internal origin. 
Sometimes the close location of the absorption system
leads to incomplete coverage of the cental light source. 
This results in flat line profiles not going to zero intensities and/or in
unexpected doublet ratios. None of these features is observed in the present
absorption systems.

Another evidence of the high activity of CTQ~325 can be found from the
distribution of metal absorbers along the line of sight (see Table~1). 
Thus, Fig.~4 
shows that in the range $3.0 \la z \la 3.1$ the number density $dN/dz$ exceeds
essentially the mean value.
However, this argument is not unambiguous since one cannot exclude the observed overdensity
as being due to the clustering of the intervening absorbers (e.g., in Fig.~4, the region
$z > 2.7$ toward Q~1157+3143 was considered as a possible supercluster
by Ganguly et al. 2001).

Although the jet-cloud origin is consistent with all recovered parameters of this
broad-hot~+~narrow-cool absorption system, another interpretation --- absorption
by warm gas in the intervening large scale structure (LLS) object --- is also 
possible (motivated mostly by the Tololo pair). 
In this case the collisionally ionized phase may be produced by
shocks driven by the accretion of gas into the potential well.
If typical size of a putative cluster is $R \sim 1$ Mpc,
then the velocity dispersion of the warm gas component of $\sim 400$ \kms\, 
gives the total mass
$M_{\rm cl} \sim 2\times10^{13}M_\odot$. 
High metallicity estimated
for the collisionally ionized gas is in line with that usually measured 
in clusters of galaxies (e.g., Rosati et al. 2002). 
The presence of the compact cool absorbers may also be expected: 
the ionic composition of our narrow line absorber at $v \simeq -150$ \kms\,
is very much alike to that observed in a cloud in the Virgo supercluster, 
(see Fig.~2 in Tripp et al. 2002). The size of this cloud was estimated as
70 pc only.
The existence of massive megaparsec-scale structures
at high redshift (from $z = 2.16$ up to $z = 4.10$) has recently been confirmed by direct
observations of distant (proto) clusters (Venemans et al. 2002, 2003).

\section{Conclusions}

We show that the shallow and broad \ion{C}{iv}, \ion{O}{vi}, and Ly$\alpha$
lines detected at \zabs = 3.021 in the spectrum of CTQ~325 are produced by
{\it collisionally} ionized gas. The narrow line absorber detected at the same
redshift reveals the ionization state expected for a cool photoionized medium~---
strong saturated hydrogen lines, absorptions in \ion{C}{iii} and \ion{Si}{iii},
and no high ionization lines like \ion{C}{iv} or \ion{O}{vi}.
Collisionally ionized gas has the near solar metallicity, whereas the metal
content of the photoionized component is less than $0.1Z_\odot$.

The observed multiphase structure can be well described in the framework of
the jet-cloud interaction. However, the absorption by the  shocked accreting
gas in the intervening LSS object cannot be excluded. At the moment there are
no unambiguous tests to select between these possibilities.

\begin{acknowledgements}
The work of S.A.L. and I.I.A. is partly supported by
the RFBR grant No. 03-02-17522.
\end{acknowledgements}

\end{document}